\documentclass{article}
\usepackage{graphicx}
\usepackage{bm}
\begin{document}

\title{Dynamics of Vortices and Solitons in a Bose-Einstein Condensate by an Oscillating Potential}
\author{Kazuya Fujimoto, Makoto Tsubota}

\maketitle

\begin{abstract}
We study numerically the dynamics of quantized vortices and solitons induced by an oscillating potential inside 
a trapped Bose-Einstein condensate. The dynamics of the topological defects 
is much different from the case for a linear uniform object; the metamorphosis between vortices and solitons 
is characteristic of the dynamics. We discuss how vortices are nucleated by an oscillating potential.


\end{abstract}

\section{Introduction}

One of the most difficult problems in classical physics is turbulence.
It is generally considered that turbulence consists of vortices. The circulation of vortices has an arbitrary value, and 
the vortices repeat nucleation and annihilation. There is a similar phenomenon in a 
quantum fluid, which is called quantum turbulence\hspace{0.5mm}(QT). In contrast to classical turbulence\hspace{0.5mm}(CT), the circulation of vortices in QT 
takes a discrete value, and quantized vortices are more stable because they are topological defects \cite{Donnely}. 
Thus the constituent elements of QT are more definite than those of CT. The Kolmogorov law, which is the 
most important statistical law in turbulence, is confirmed in QT numerically with the Gross-Pitaevskii\hspace{0.5mm}(GP) model\cite{kolomogrov}; these observations 
show some significant similarity between CT and QT in spite of the difference of the nature of vortices.

A soliton is also important for quantum fluids, and there exists much literature on the subject. 
Solitons and vortices are the representative topological defects in quantum fluids. Then it would be significant to study the synergy dynamics between 
vortices and solitons.

Quantum turbulence has been studied in superfluid $^4$He and $^3$He for these years \cite{turbulence}. Recently, QT is realized also in an 
atomic Bose-Einstein condensate\hspace{0.5mm}(BEC) \cite{turbulence experiment}. The key word for common technique making QT experimentally is oscillation. 
For example, QT in He is realized by oscillating the objects such as wires, grids, spheres and tuning folks, while  oscillating 
a trapping potential leads to QT in atomic BECs. However, there are not so many 
works with oscillating potential inside trapped BECs\cite{oscillation,k}.   

In this paper, we study synergy dynamics of solitons and vortices induced by an oscillating potential inside a trapped BECs, and 
discuss the vortex nucleation characterized by oscillation. This approach should open a new aspect of quantum hydrodynamics.

\section{Formulation}
 We consider a dilute atomic gas BEC with a macroscopic wave function, assuming that the condensate is pancake shaped. Then 
the macroscopic wave function $ \psi(\bm{x},t) $ obeys the two-dimensional Gross-Pitaevskii equation:
\begin{equation}
  i\hbar\frac{\partial  }{\partial t} \psi(  \bm{x},t) =- \frac{ \hbar^2} {2m} \triangle \psi(  \bm{x},t)+V( \bm{x},t)\psi(  \bm{x},t) + g| \psi(  \bm{x},t)|^2 \psi(  \bm{x},t),
\end{equation}
where $ m $ is a particle mass, $ g $ is an interaction parameter for two-dimensional case, and $ V( \bm{x},t) $ is potential. The macroscopic wave
function $ \psi(\bm{x},t) $ is normalized by the total particle number $ N $.
We suppose that the condensate is trapped by a harmonic potential and penetrated by a Gaussian potential, so
$ V( \bm{x},t) = V_{\rm trap}( \bm{x}) + V_{\rm object}( \bm{x},t) $. Here $ V_{\rm trap}( \bm{x}) $ is harmonic as usual, 
$ V_{\rm trap}( \bm{x})= \frac{1}{2}m(\omega^2_xx^2+\omega^2_yy^2 ) $, and $ V_{\rm object}( \bm{x},t) $ is represented by 
\begin{equation}
V_{\rm object}( \bm{r},t) = V_0{\rm exp}[-((x-x_0(t))^2+y^2)/d^2],
\end{equation}
where $ x_0(t) $ is the $x$-coordinate of the center of the potential and $ d $ is the radius of the Gaussian potential. 
We move the Gaussian potential like $ x_0(t)=\epsilon \hspace{0.5mm} \rm{sin( \omega \it t)} $ with the oscillating amplitude $\epsilon$ and 
the frequency $\omega$. The velocity of the potential is $v_{\rm p}=\epsilon \omega$.

We numerically solve Eq.\hspace{0.5mm}(1) to 
investigate the dynamics of the topological defects\hspace{0.5mm}(vortices and solitons). 
Neither dissipation nor noise is included in our calculations.

\section{Numerical results}

We choose the parameters as $ g=4.19 \times 10^{-45} \hspace{0.5mm}\rm{J/m^2} $, $ m=1.42 \times 10^{25} \hspace{0.5mm}\rm{kg} $, $ N=6.6 \times 10^{4} $, $ \omega _x= 2 \pi \times 5 $\hspace{0.5mm}/s, 
$ \omega _y= 2 \pi \times 25 $\hspace{0.5mm}/s, $d=6 \hspace{0.5mm}\mu \rm m$, $\epsilon =7 \hspace{0.5mm}\rm{\mu} m$, $\omega=100$\hspace{0.5mm}/s and $ V_0= 60gn_0 $, where $n_0$ is the density near the center of the condensate.  
In these parameters the Thomas-Fermi radii are $R_x=59.6 \hspace{0.5mm}\mu \rm m$ and $R_y=11.9 \hspace{0.5mm}\mu \rm m$, and the sound speed near the center of the condensate is $ 1.33
\times 10^{3} \hspace{0.5mm}\mu \rm{m/s} $.
We start the simulation with an initial stationary state of Fig.\hspace{0.5mm}1 obtained by the imaginary time step of the GP equation.

This section consists of four parts. We describe elementary dynamics for vortices and solitons in the first three parts and show the global dynamics in the last part.
One of characteristics of dynamics is metamorphosis between vortices and solitons. The solitons appearing in our calculations are related with 
rarefaction pulse \cite{rare1, rare2} which is finite 
amplitude sound wave.

Here we discuss heating in this system because the oscillating potential may increase the temperature and break a BEC. 
We can numerically calculate the increase of the total energy, 
estimating a change of temperature by using specific heat for equilibrium state. 
The increase of the temperature is found to be the order of 10\hspace{0.5mm}nK. 
Hence the heating can be ignored, and our numerical calculations with the GP model are valid.

\subsection*{Nucleation of a vortex pair}

\begin{figure}[t]
\begin{center}
\includegraphics[keepaspectratio, width=9cm]{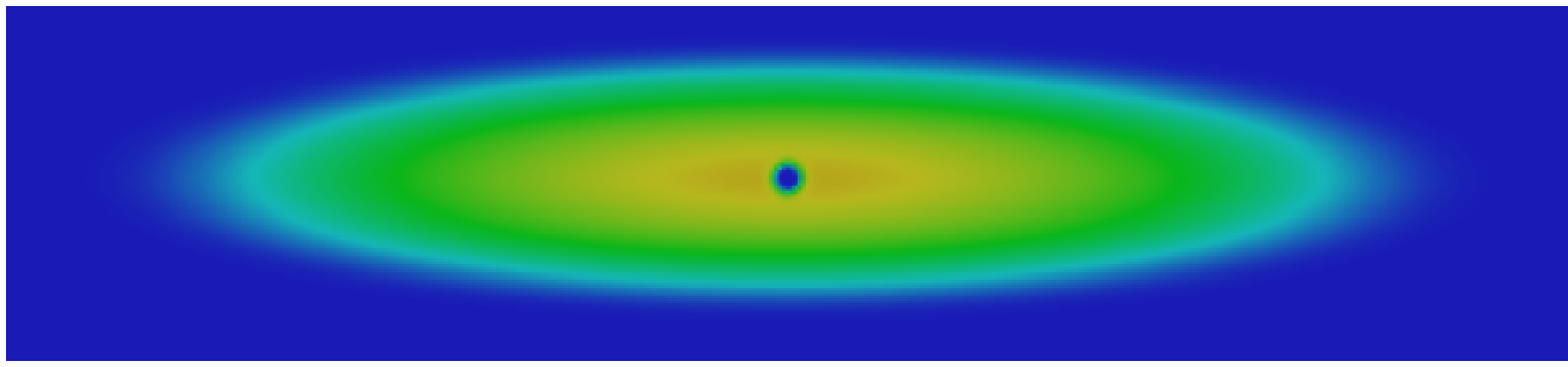}
\includegraphics[width=0.1cm, height=1.5cm]{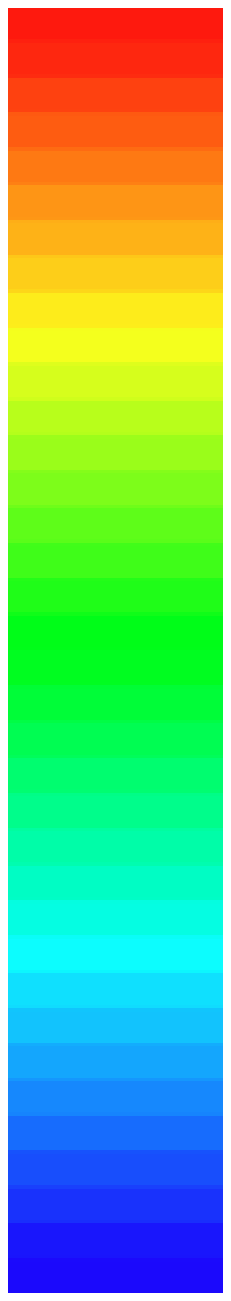}
\begin{picture}(10,40)(0,0)
\put(0,0){0}
\put(0,40){$1.8 \times 10^{-4}$}
\put(8, 0){\vector(0, 1){30}}
\put(8, 0){\vector(1, 0){30}}
\put(34,3){$x$}
\put(11,26){$y$}
\end{picture}
\end{center}
\caption{(color online) Initial state of the condensate density: The $ x $ and $ y $ sizes of the 
box are $140 \hspace{0.5mm}\rm{\mu m}$ and $33 \hspace{0.5mm}\rm{\mu m}$. 
Thomas-Fermi radii are $R_x=59.6 \hspace{0.5mm}\mu \rm{m}$ and $R_y=11.9 \hspace{0.5mm}\mu \rm{m}$. The low density region in the center comes from 
the Gaussian potential with the radius $6\hspace{0.5mm} \mu \rm{m} $. The right bar shows the range of the dimensionless density. This bar is common through 
Figs\hspace{0.5mm}.$1\sim 6$.}

\end{figure}

\begin{figure}[t]
\begin{center}
\begin{picture}(200,140)(0,0)
\put(-35,0){\includegraphics[keepaspectratio, width=9cm]{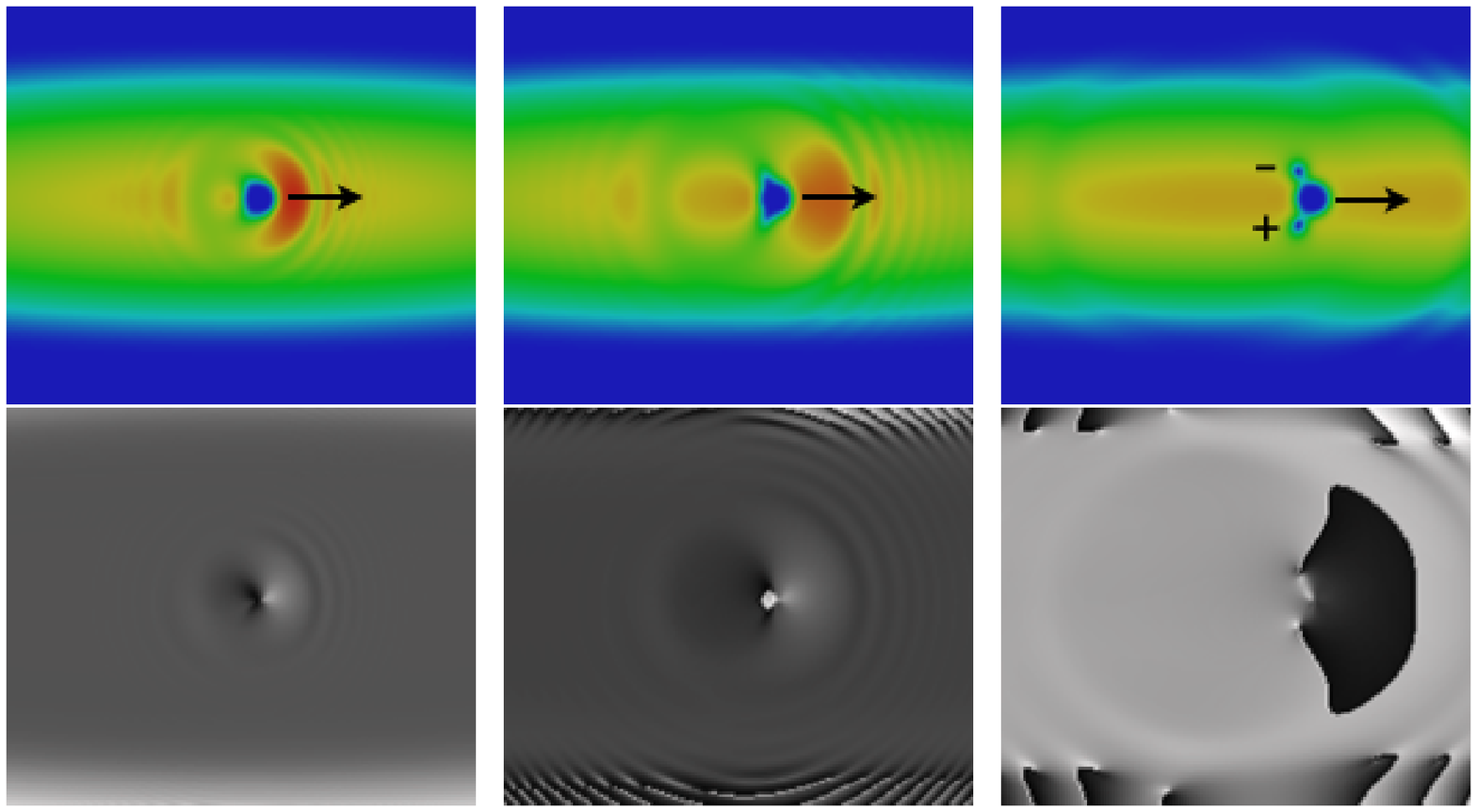}}
\put(225,0){\includegraphics[width=0.1cm, height=1.5cm]{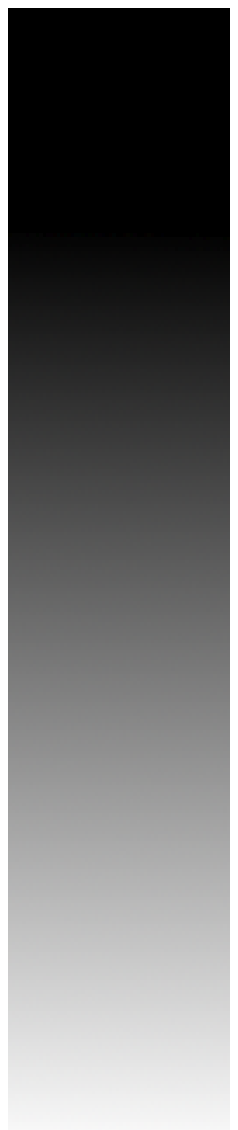}}
\put(225,70.6){\includegraphics[width=0.1cm, height=1.5cm]{bar.eps}}
\put(-45,95){\rotatebox{90}{density}}
\put(-45,25){\rotatebox{90}{phase}}
\put(230,0){$-\pi $}
\put(230,40){$ \pi $}
\put(230,70.6){0}
\put(230,110.6){$1.8 \times 10^{-4}$}
\put(-32, 3){\vector(0, 1){30}}
\put(-32, 3){\vector(1, 0){30}}
\put(-7,6){$x$}
\put(-30,29){$y$}
\put(-32, 75){\vector(0, 1){30}}
\put(-32, 75){\vector(1, 0){30}}
\put(-7,82){$x$}
\put(-30,111){$y$}
\end{picture}
\end{center}
\hspace{2.35cm}(a)\hspace{2.8cm}(b)\hspace{2.8cm}(c)
\caption{(color online) Vortex nucleation: (a) $t$=0.00253\hspace{0.5mm}s, (b) $t$=0.00490\hspace{0.5mm}s, (c) $t$=0.0118\hspace{0.5mm}s. 
The upper and lower pictures denote the density and phase profiles. 
The $x$ and $y$ sizes of box are $36 \hspace{0.5mm}\rm{\mu m}$ and $33 \hspace{0.5mm}\rm{\mu m}$ through Figs\hspace{0.5mm}.$2\sim 5$. 
The symbol + denotes a vortex with the clockwise circulation and - denotes a vortex with the counter-clockwise circulation. 
Arrows refer to the direction of velocity of the potential.}

\end{figure}

We discuss the nucleation of a vortex pair. Two kinds of vortices appear in our numerical calculations. One is a usual vortex which wears the profile 
of the condensate density, and the other is a ghost vortex\cite{ghost} which exists in the very low density region and cannot be observed by the density measurement. 
We consider nucleation of $both$ kinds of vortices in this work. Then there are two critical velocities; $v_{\rm c1}$ for a pair of ghost vortices and $v_{\rm c2}$ 
for a pair of vortices with $ v_{\rm c1}<v_{\rm c2} $. When the velocity $v_{\rm p}$ of the potential is smaller than $v_{\rm c1}$, any vortices are not nucleated in 
the condensate. As $v_{\rm p}$ is increased to exceed $v_{\rm c1}$, a pair of ghost vortices is nucleated inside the Gaussian potential. If $v_{\rm p}$ is smaller than 
$v_{\rm c2}$, however, ghost vortices are immediately annihilated. Thus ghost vortices repeat nucleation and annihilation inside the potential, so that 
they cannot get away from it and vortices are never nucleated in the condensate. When $v_{\rm p}$ exceeds $v_{\rm c2}$, the vortices are nucleated eventually. 

This process is shown in Fig.\hspace{0.5mm}2. Firstly, the velocity field like back-flow is induced by the potential in Fig.\hspace{0.5mm}2(a). Secondly, a pair 
of ghost vortices is nucleated inside the potential in Fig.\hspace{0.5mm}2(b). Thirdly, the ghost vortices get away from it, then a vortex pair is nucleated 
in Fig.\hspace{0.5mm}2(c). The repeated nucleation and annihilation of a pair of ghost vortices are characteristic 
of the oscillating case with $v_{\rm c1} < v_{\rm p} < v_{\rm c2}$. In the present case, $v_{\rm c1} \cong 500\hspace{0.5mm} \rm{\mu m/s}$ and 
$v_{\rm c2} \cong 580\hspace{0.5mm} \rm{\mu m/s}$, but $v_{\rm c1}$
and $v_{\rm c2}$ strongly depend on the amplitude $\epsilon$. The critical velocity $v_{\rm c2}$ is smaller than the sound speed $1.33
\times 10^{3}$\hspace{0.5mm}$\rm \mu m/s$, 
similar things are reported by other authors. The dependence should be systematically investigated, reported soon elsewhere.

\subsection*{Dynamics of vortices}

\begin{figure}[t]
\begin{center}
\includegraphics[keepaspectratio,width=11cm]{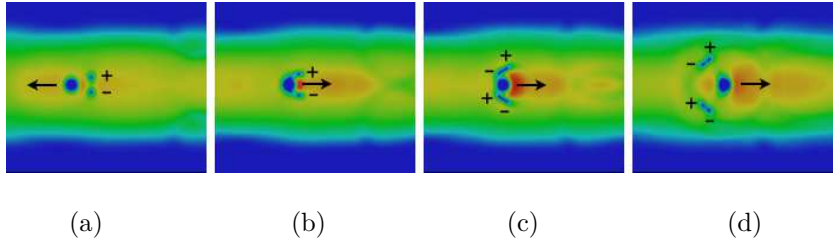}
\end{center}
\hspace{1.4cm}(a)\hspace{2.5cm}(b)\hspace{2.4cm}(c)\hspace{2.5cm}(d)
\caption{(color online) Dynamics of vortices: (a) $t$=0.0507\hspace{0.5mm}s, (b) $t$=0.0551\hspace{0.5mm}s, 
(c) $t$=0.0571\hspace{0.5mm}s, (d) $t$=0.0596\hspace{0.5mm}s. These pictures denotes the density profile.
The symbol + denotes a vortex with the clockwise circulation and - denotes a vortex with the counter-clockwise circulation. 
Arrows refer to the direction of velocity of the potential.}

\end{figure}

In this subsection, we show the dynamics of vortices in Fig.\hspace{0.5mm}3, immediately after the potential starts to oscillate.
The dynamics of vortices by an oscillating potential is quite different from that by a linear uniform moving potential.
Figure\hspace{0.5mm}3 shows the characteristic dynamics of vortices. As the potential moves, a vortex pair is nucleated behind 
the potential through the process in Fig.\hspace{0.5mm}2, and has the impulse whose direction is the same as the velocity of the potential in Fig.\hspace{0.5mm}3(a). 
Consequently, the pair follows the potential. 
After the oscillating potential changes the direction in Fig.\hspace{0.5mm}3(b), another pair of ghost vortices 
is nucleated inside the potential and the old pair collides with the potential.
Through this collision, the ghost vortices come out of the potential to become a new vortex pair, which reconnects with the old pair in Fig.\hspace{0.5mm}3(c), 
then the rearranged pairs leave the potential and move towards 
the surface of the condensate in Fig.\hspace{0.5mm}3(d). This phenomenon is peculiar to the oscillating potential\hspace{0.5mm}(AC), not observed in linear 
moving potential\hspace{0.5mm}(DC).
Note that the ghost vortices are nucleated in the surface of the condensate as the potential moves.
These pairs in Fig.\hspace{0.5mm}3(d) arrive at the surface, and reconnect with ghost vortices. 
Afterward, new vortex pairs move along the surface of the condensate and head toward the bow of it\cite{so}.

\subsection*{Dynamics of solitons}

Our calculation shows the interesting and remarkably dynamics that solitons are nucleated by annihilation of vortex pairs and the collapse of solitons makes
vortex pairs.

\begin{figure}[t]
\begin{center}
\includegraphics[keepaspectratio, width=11cm]{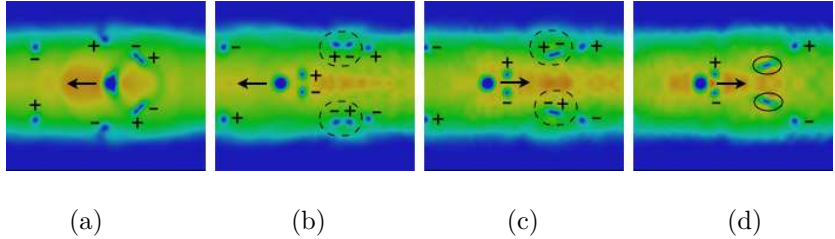} 
\end{center} 
\hspace{1.4cm}(a)\hspace{2.5cm}(b)\hspace{2.4cm}(c)\hspace{2.5cm}(d)
\caption{(color online) Nucleation of solitons: (a) $t$=0.0929\hspace{0.5mm}s, (b) $t$=0.106\hspace{0.5mm}s, (c) $t$=0.111\hspace{0.5mm}s, (d) $t$=0.114\hspace{0.5mm}s.
These pictures show the density profile. The symbol + denotes a vortex with the clockwise circulation and - denotes a vortex with the counter-clockwise circulation. 
Arrows refer to the direction of velocity of the potential. In the process from (a) to (b), reconnection of vortices occurs, then 
two new vortex pairs, which will become solitons, are created. These pairs are depicted by closed loops with broken lines. The closed loops with solid lines in (d) show the solitons.}
\end{figure}

\begin{figure}[t]
\begin{center}
\includegraphics[keepaspectratio, width=11cm]{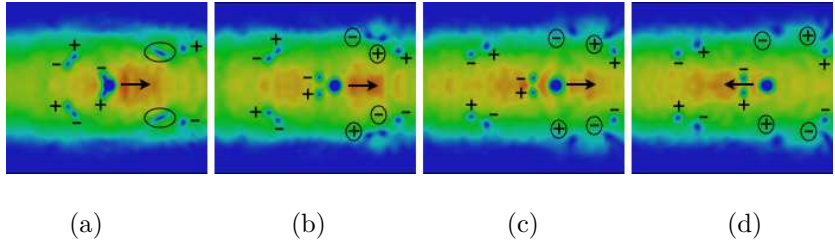}
\end{center}
\hspace{1.4cm}(a)\hspace{2.5cm}(b)\hspace{2.4cm}(c)\hspace{2.5cm}(d)
\caption{(color online) Collapse of solitons to vortex pair and their dynamics: (a) $t$=0.126\hspace{0.5mm}s, (b) $t$=0.131\hspace{0.5mm}s, 
(c) $t$=0.136\hspace{0.5mm}s, (d) $t$=0.138\hspace{0.5mm}s. 
These pictures show the density profile. The symbol + denotes a vortex with the clockwise circulation and - denotes a vortex with the counter-clockwise circulation. 
Arrows refer to the direction of velocity of the potential. The closed loops with solid lines in (a) show the solitons. 
The symbols $\oplus$ and $\ominus$ denote the vortices created 
through the collapse of the soliton.}

\end{figure}

The annihilation of vortex pairs makes solitons. 
Figure\hspace{0.5mm}4 shows this process. 
The vortex configuration in Fig.\hspace{0.5mm}4(a) is obtained by repeating the processes in Fig.\hspace{0.5mm}3 two times.
Then the vortex pairs rearrange near 
the surface in Fig.\hspace{0.5mm}4(b) and new vortex pairs move to the central part of the condensate in Fig.\hspace{0.5mm}4(c). 
These new vortex pairs are depicted by closed loops with broken lines.
Then the annihilation of the vortex pairs nucleates solitons in Fig.\hspace{0.5mm}4(d).
The soliton created by the annihilation are related to rarefaction pulse.

Next we show that the collapse of the solitons leads to the nucleation of another vortex pair. 
This process occurs after Fig.\hspace{0.5mm}4, shown in Fig.\hspace{0.5mm}5.
After Fig.\hspace{0.5mm}4(d), the solitons go through each other without changing the shape, which is characteristic of soliton. 
Then Fig.\hspace{0.5mm}5(a) is obtained. 
The solitons go toward the surface in Fig.\hspace{0.5mm}5(a) and the solitons collapse to vortex pairs\cite{na} near the surface in Fig.\hspace{0.5mm}5(b). 
The reason why this collapse occurs in the surface may be that the phase of the macroscopic wave function is disturbed in the low density region. 
Later, the new vortex pairs reconnect with the ghost vortices in Fig.\hspace{0.5mm}5(c) and the new pairs head toward the bow of the condensate in Fig.\hspace{0.5mm}5(d).

\subsection*{Global dynamics of vortices}

\begin{figure}[t]
\begin{center}
\includegraphics[keepaspectratio, width=11cm]{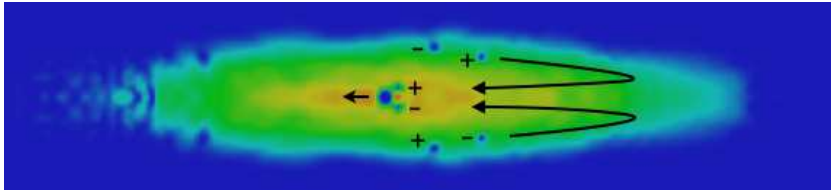}
\end{center}
\caption{(color online) Trajectories of vortices: The $ x $ and $ y $ sizes of the 
box are $140 \hspace{0.5mm}\rm{\mu m}$ and $33 \hspace{0.5mm}\rm{\mu m}$. The snapshot is the density profile at $t$=0.0845\hspace{0.5mm}s. The symbol + denotes a vortex with the clockwise circulation and - denotes a vortex with the counter-clockwise circulation. Arrows refer to the direction of velocity of the potential.The curves show the trajectories of vortices which do not 
change to the solitons. The same motion occurs in the left side of the condensate.}

\end{figure}

Global dynamics of vortices consists of the above elementary processes. 
In Fig.\hspace{0.5mm}6 we confine our description to right region of the condensate.
The vortices nucleated by the Gaussian potential leave it by the process shown in Fig.\hspace{0.5mm}3. 
Subsequently, the vortices reach the surface and reconnect with 
ghost vortices. The resulting pairs at the upper and lower sides of the condensate go to the bow along the surface, then the pairs meet near the bow, and again reconnect 
with each other. As a result appear a pair of vortices which heads toward the center of the condensate.

\section{Conclusion}
We perform the numerical research with the GP model to investigate the dynamics and nucleation of vortices and solitons induced by an oscillating potential 
in a trapped BEC. Our calculations found three interesting phenomena. Firstly, the dynamics of AC case is much different from that of DC case in many respects, 
while they are simmilar in some respects. For example, 
the trajectories of the vortices shown in Fig.\hspace{0.5mm}6 are observed in a DC experiment\cite{anderson}, but the dynamics shown in Fig.\hspace{0.5mm}3 is 
characteristic of the AC case. 
Secondly, the dynamics and nucleation of vortices are deeply involved with the ghost vortices. In particular, we consider that the ghost vortices may be essential 
to the nucleation for the vortices. Thirdly, the synergy dynamics between vortices and soliton is realized. This is related with the rarefaction pulse. 
One can enjoy the movie of the numerical simulation in our web site\cite{hp}. 

The parameters used in our calculations are realistic in experiments of atomic BECs. Hence, the dynamics obtained by our study can be observed actually 
in experiments. We consider that oscillating Gaussian potential may lead to QT so that this method may be a new technique for making QT in atomic BECs. 
The details reviewed will be reported soon elsewhere.

\end{document}